\documentclass[aps,prb,twocolumn,showpacs,floatfix,superscriptaddress,nofootinbib]{revtex4}
\usepackage{graphicx}
\usepackage{amssymb}
\usepackage{amsmath}
\usepackage{lmodern}
\usepackage{color}
\usepackage{hyperref}
\usepackage{empheq}
\usepackage{epstopdf}
\usepackage{amsmath}

\DeclareMathOperator{\sech}{sech}

\begin{document}

\title{Exact analytical solutions for the kinks, the solitons and the shocks in  discrete nonlinear transmission line with nonlinear capacitance}

\author{Eugene Kogan}
\email{Eugene.Kogan@biu.ac.il}
\affiliation{Department of Physics, Bar-Ilan University, Ramat-Gan 52900, Israel}

\begin{abstract}

We studied  discrete transmission lines constructed from  ideal linear inductors and nonlinear  capacitors (and possibly resistors). The  localised travelling waves  in the lossless transmission lines are the kinks and the solitons, the  localised travelling waves  in the lossy transmission lines
are the dissipative kinks and the shocks. The speeds and  profiles of all these waves were  calculated.

\end{abstract}

\date{\today}

\maketitle

\section{Introduction}

The  nonlinear electrical transmission lines are of much interest both due to their applications, and as the laboratories to study nonlinear waves \cite{malomed2}.
Among the latter -- the  travelling waves, especially the kinks and the solitons, attract a lot of attention \cite{solitons}. The nonlinear waves become especially interesting (and challenging) when one takes into account the discrete nature of the transmission line. The kinks and the solitons in  discrete transmission lines were studied in quite a few publications \cite{malomed4,cameroon,bogdan}.

The field of  nonlinear waves in discrete transmission lines is closely connected with Fermi-Pasta-Ulam-Tsingou (FPUT) problem \cite{fermi}. In particular, dynamics of lattice kinks and solitons were studied  in a large number of publications \cite{kevrekidis,vainchtein,vainchtein2,kamchatnov}

We studied previously the  kinks and the solitons in discrete Josephson transmission line (JTL) JTL \cite{kogan2,kogan3}. There we formulated the quasi-static approximation, which allows to reduce the discrete equations to the continuous ones. In the present short note we would like to apply this approximation to different kind of discrete transmission lines -- with nonlinear  capacitors. Such transmission lines in the continuum approximation were studied in the classical papers by R. Landauer \cite{landauer1,landauer2,landauer}.

\section{The kinks and the solitons}
\label{con}

\subsection{The travelling waves}
\label{con2}

\begin{figure}[h]
\includegraphics[width=\columnwidth]{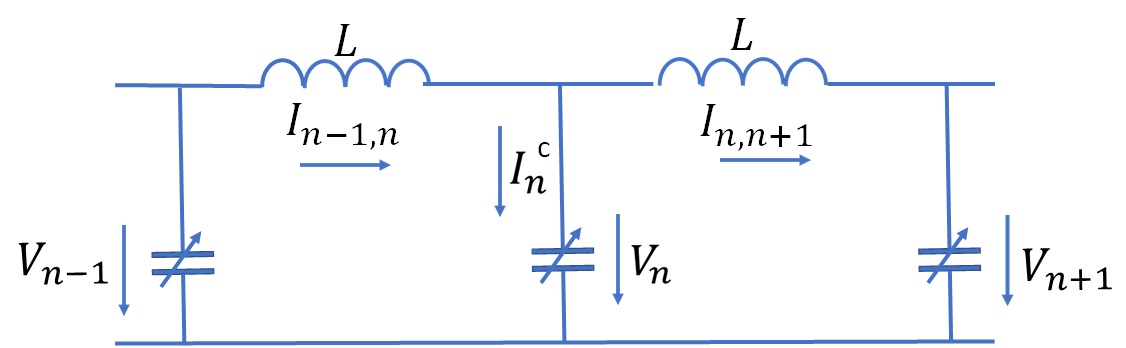}
\caption{The lossless discrete  transmission line.}
 \label{trans5}
\end{figure}

The discrete  transmission line is constructed from the identical inductors   and nonlinear capacitors is shown on Fig. \ref{trans5}.

We take the voltages $V_n$ and the currents which pass though the inductors $I_{n-1,n}$ as the dynamical variables.
The circuit equations (Kirchhoff laws) are
\begin{subequations}
\label{a10}
\begin{alignat}{4}
\frac{dQ_n}{dt}&=  I^c_n,\label{a8b}\\
L\frac{d I_{n,n+1}}{d t}&=V_{n}-V_{n+1}\label{8a},\\
I^c_n&=I_{n-1,n}-I_{n,n+1}.\label{kl}
\end{alignat}
\end{subequations}
where  $Q_n=Q(V_n)$ and the nonlinear nature of the transmission line is determined by the (monotonically increasing) function  $Q=Q(V)$.
We can exclude $I_{n,n+1}$ from (\ref{8a}) and (\ref{kl}) and present these two equations a single one
\begin{eqnarray}
\label{a8a}
L\frac{d I^c_n}{d t}=V_{n+1}-2V_{n}+V_{n-1}.
\end{eqnarray}
From (\ref{a8b}) and (\ref{a8a}) we can obtain closed wave equation for the voltage $V$
\begin{eqnarray}
\label{comic}
L\frac{d^2 Q_n}{d t^2}=V_{n+1}-2V_{n}+V_{n-1}.
\end{eqnarray}

Let us write down Eq. (\ref{comic}) in the quasi-continuum approximation,
which was proposed in our previous publications \cite{kogan2,kogan3}.
The approximation consists in treating $n$ as the continuous variable $Z $, expanding
 the r.h.s. of the equation (\ref{comic}) in Taylor series with respect to $Z$ and keeping only the first two terms of the expansion, thus obtaining the equation in the form
\begin{eqnarray}
\label{com}
L\frac{\partial^2 Q(V)}{\partial t^2}=
\frac{\partial^2 V}{\partial Z^2}+\frac{1}{12}
\frac{\partial^4V}{\partial Z^4}.
\end{eqnarray}

In this paper we are interested in the
travelling wave solutions, for which the dependence of all the variables upon $Z$ and $t$ is of the form
\begin{eqnarray}
\label{comsi}
X(Z,t)=X(Z+Ut),
\end{eqnarray}
where $U$ is the speed of the travelling wave (the distance is measured in the units of the transmission line spacial period $\Lambda$). This from (\ref{com}) we obtain
an ordinary differential equation
\begin{eqnarray}
\label{combo}
U^2L\frac{d^2 Q(V)}{d Z^2}=
\frac{d^2 V}{d Z^2}+\frac{1}{12}\frac{d^4V}{d Z^4}.
\end{eqnarray}
After finding $V(Z)$
we can from (\ref{8a})  obtain the current
\begin{eqnarray}
\label{i2}
I(Z)=-\frac{V(Z)}{UL}+\text{const}.
\end{eqnarray}

Performing an integration  in (\ref{combo}) we can get
\begin{eqnarray}
\label{new4}
\frac{1}{12}\frac{d^3V}{d Z^3}=U^2L\frac{dQ}{d Z} -\frac{dV}{d Z}.
\end{eqnarray}
Notice that equation with the same form as (\ref{new4}) one would obtain after making the travelling wave ansatz in the generalised  Korteweg-de Vries-modified Korteweg-de Vries (KDV-mKDV) equation.

We may perform the second integration to obtain
\begin{eqnarray}
\label{n45bu}
\frac{1}{12}\frac{d^2V}{d Z^2}=U^2LQ(V)-V+F,
\end{eqnarray}
where $F$ is the constant of integration.
Introducing the new dependent variable
\begin{eqnarray}
\label{def}
E= \frac{dV}{dZ},
\end{eqnarray}
considering $V$ as the independent variable and
taking into account that
\begin{eqnarray}
\label{ty}
\frac{d^2V}{dZ^2}=\frac{1}{2}\frac{dE^2}{dV},
\end{eqnarray}
we may present  (\ref{n45bu}) as
\begin{eqnarray}
\frac{1}{24}\frac{dE^2}{d V}=U^2LQ(V)-V+F.
\end{eqnarray}

However, for technical reasons, it will be more convenient to return to
(\ref{new4}) and  avoid the integration of the equation.
After presenting $dQ/dZ$ as  $(dQ/dV)(dV/dZ)$, taking into account that
\begin{eqnarray}
\frac{d^3V}{dZ^3}=\frac{1}{2}E\frac{d^2E^2}{dV^2},
\end{eqnarray}
and dividing both parts of  (\ref{new4}) by the common multiplier $E$ in both parts,
we may reduce the order  of the equation and write it down as
\begin{eqnarray}
\label{22}
\frac{1}{24}\frac{d^2E^2}{dV^2}=U^2L\frac{dQ(V)}{d V}-1.
\end{eqnarray}

Consider the "sound" waves that is  small amplitude, smooth waves propagating on the homogeneous backgrounds $V$. These waves are described by Eq. (\ref{22}), with the l.h.s. being put equal to zero.
The speed of the sound is given by the equation
\begin{eqnarray}
\label{velocity}
\frac{1}{u^2(V)}=L\frac{dQ}{dV}.
\end{eqnarray}
Thus Eq. (\ref{22}) can be presented as
\begin{eqnarray}
\label{22g}
\frac{1}{24}\frac{d^2E^2}{dV^2}=\frac{U^2}{u^2(V)}-1.
\end{eqnarray}

Note that in the framework of the quasi-continuum approximation $E$ is just the voltage on the inductor. Thus the physical meaning of Eq. (\ref{22g}) is simple.
Its solution gives the voltage on the inductor as the function of the voltage on the capacitor.

A simple mechanical interpretation of Eq. (\ref{22g}) may be helpful.
If we call $V$ the "time", the equation can be considered as describing motion of the  fictitious Newtonian particle, $E^2$  being its coordinate  and $1/24$ it's mass, under the influence of the time-dependent force, given by the r.h.s. of the equation.

We'll  consider  localised travelling waves, that is  impose upon $V(Z)$ the boundary conditions
\begin{eqnarray}
\label{vk}
\lim_{Z\to-\infty}V(Z)=V_1,\hskip 1cm \lim_{Z\to+\infty}V(Z)=V_2.
\end{eqnarray}
The  value of $V_2$ may be    different from $V_1$, in which case we'll talk about the kink, or equal to $V_1$, in which case we'll talk about the soliton.
The quantity $E^2(V)$ obviously has minimum (equal to zero) at $V=V_{1,2}$. Hence
from (\ref{22g}) follows
\begin{eqnarray}
U^2>u^2\left(V_{1,2}\right),
\end{eqnarray}
that is the travelling waves  are supersonic \cite{kogan2,kogan3}.

The boundary conditions (\ref{vk}) obviously give
\begin{eqnarray}
\label{vk7}
E\left(V_{1,2}\right)=0.
\end{eqnarray}
Also, from (\ref{ty}) and (\ref{vk}) follows that in the vicinity of $V_{1,2}$
\begin{eqnarray}
\label{vi}
E^2\sim \left(V-V_{1,2}\right)^2.
\end{eqnarray}
The conditions
(\ref{vk7}) and (\ref{vi})  mean that the particle is at rest at the origin at "time" equal to $V_1$ and then again at "time" equal to $V_2$.

The main result of the Section \ref{con2} is Eq. (\ref{22}).
The natural question arises: why we were not happy with Eq. (\ref{n45bu}) (which can be easily integrated in quadratures) and introduced instead of it the former equation. The detailed answer to this question will be given in Section  \ref{lossy},
but we can formulate it announce it right now. The main aim of the present paper is obtaining exact analytical solutions for the kinks and the shocks in the lossy transmission line. When the loss terms are included into the equations, (\ref{22})
can be integrated analytically much easier than (\ref{n45bu}).

\subsection{The kinks}
\label{kinik}

With four boundary conditions specifying both the position and the velocity of the particle at $V=V_1$ and $V=V_2$,  of the three parameters $V_1,V_2,U^2$ only one (say $V_1$) can be considered as a free parameter, the other two are connected with it.
One such constraint we could obtain at much earlier stage. By integrating (\ref{22}) with respect to $V$ from $V_1$ to $V_2$ and taking into account the boundary conditions we obtain the speed of the kink
\begin{eqnarray}
\label{ququ}
\frac{1}{U^2}=L\frac{Q_2-Q_1}{V_2-V_1}.
\end{eqnarray}
where $Q_i=Q(V_i)$.

If $Q(V)$ is a polynomial function of $V$,
from  (\ref{vi}) follows that the kink  can be presented as
\begin{eqnarray}
\label{ee}
E^2(V)=P_1(V)(V-V_1)^2(V-V_2)^2,
\end{eqnarray}
where $P_1(V)$ is some polynomial function.

If we  approximate $Q(V)$ between $V_1$ and $V_2$ by the third degree polynomial
\begin{eqnarray}
\label{delo}
Q(V)=\text{const}+C_0\left(\alpha V+\beta  V^2+\gamma V^3\right)
\end{eqnarray}
(for brevity we'll call Eq. (\ref{delo})  the cubic case),
then
\begin{subequations}
\label{speeds}
\begin{align}
\frac{1}{u^2(V)}&= LC_0\left(\alpha+2\beta V+3\gamma V^2\right),  \label{sound}\\
\frac{1}{U^2}&=LC_0\left[\alpha+\beta\left(V_1+V_2\right)+
\gamma\left(V_1^2+V_1V_2+V_2^2\right)\right].\label{qu}
\end{align}
\end{subequations}
We also can specify (\ref{ee}) to
\begin{eqnarray}
\label{eem}
E(V)=\psi(V-V_1)(V-V_2),
\end{eqnarray}
where $\psi$ is a constant (positive, if $V_1>V_2$, and negative in the opposite case).
Substituting (\ref{eem})  into (\ref{22g}) and using Vieta's formula  we  obtain
after simple algebra
\begin{subequations}
\label{coc}
\begin{alignat}{4}
\psi^2&=6U^2LC_0\gamma, \label{velos}\\
V_1+V_2&=-\frac{2\beta}{3\gamma}. \label{k2}
\end{alignat}
\end{subequations}
Thus the necessary condition for the existence of the kink is the inequality $\gamma>0$, that is the speed of sound $u^2(V)$ should be a convex function of $V$.

To find $V(Z)$ we have to solve Eq. (\ref{def}).
The solution   is
\begin{eqnarray}
\label{kk}
V(Z)=\frac{V_1+V_2}{2}-\frac{V_1-V_2}{2}
\tanh \left[\frac{V_1-V_2}{2}\psi  Z\right].
\end{eqnarray}

\subsection{The  solitons}
\label{solist}

To specify the soliton we need to introduce
an additional parameter, and as such we'll take the extremal value of the voltage $V_3$.
That is $V$ changes between $V_1$ and $V_3$, and $|V_3-V_1|$ we'll call the amplitude of the soliton.
Because  $V_3$ is the extremal value of $V$ we have
\begin{eqnarray}
\label{vis}
\left(\frac{dV}{dZ}\right)_{V=V_3}=0, \hskip 1cm \left(\frac{d^2V}{dZ^2}\right)_{V=V_3}\neq 0.
\end{eqnarray}
Hence from (\ref{def}) and (\ref{ty}) follows that  in the vicinity of  $V_{3}$
\begin{eqnarray}
\label{vii}
E^2\sim \left(V-V_3\right).
\end{eqnarray}

Recalling our mechanical analogy we see that the soliton corresponds to the motion of the fictitious particle with zero initial coordinate and velocity at $V=V_1$ and zero coordinate (but not velocity) at $V=V_3$.
Mechanical interpretation of (\ref{22}) in the case of the soliton leads to the  situation which doesn't normally occur in mechanics.
The boundary condition $V_2=V_1$ means that the "time" $V$ flows along the contour from $V_1$ to $V_3$ and then back from $V_3$ to $V_1$. This is why $E(V)$  for the soliton will later turn out to be a double-valued function.

With the position and the velocity of the particle at  $V=V_1$ and its position at $V=V_3$ being specified,    of the three parameters $V_1, V_3,U^2$ only two (say $V_1$ and $V_3$) can be considered as  free parameters.
Multiplying both sides of (\ref{new4})  by $V_3-V$, integrating the equation with respect to $V$ from $V_1$ to $V_3$ and taking into account the boundary conditions we obtain the speed of the soliton
\begin{eqnarray}
\label{v5}
\frac{1}{U^2}=\frac{2}{\left(V_3-V_1\right)^2}\int_{V_1}^{V_3}\frac{V_3-V}{u^2(V)}
dV.
\end{eqnarray}

If $Q(V)$ is a polynomial function of $V$,
 from  (\ref{vi}) and (\ref{vii}) follows that the soliton can be presented as
\begin{eqnarray}
\label{ee4}
E^2(V)=P_2(V)(V-V_1)^2(V-V_3),
\end{eqnarray}
where  $P_2(V)$  is some polynomial function.

In the cubic case, calculating the integral in (\ref{v5}) we obtain
\begin{eqnarray}
\label{dd}
\frac{1}{U^2}=LC_0\left[\alpha+\frac{2\beta}{3}\left(2V_1
+V_3\right)+\frac{\gamma}{2}\left(3V_1^2+2V_1V_3+V_3^2\right)\right].
\nonumber\\
\end{eqnarray}
We  also can specify (\ref{ee4})  to
\begin{eqnarray}
\label{eem2}
E^2(V)=\psi(V-V_1)^2(V-V_3)(V-V_4),
\end{eqnarray}where $\psi$ is a constant.
Substituting (\ref{eem2})  into (\ref{22g}) and using Vieta's formula  we  obtain
after simple algebra
\begin{subequations}
\label{wu}
\begin{alignat}{4}
&\psi=6U^2LC_0\gamma, \label{lo22}\\
&2V_1+V_3+V_4=-\frac{4\beta}{3\gamma}. \label{kp2}
\end{alignat}
\end{subequations}

To make life easier to ourselves, let us restrict the consideration by the case
$|\beta/\gamma|\gg|V_1|,|V_3|$. Physically the approximation means that $V_1$ and $V_3$ are close enough to each other, and we can ignore the cubic term in (\ref{delo}).
In this case, excluding $V_4$ and introducing the amplitude $\phi$ instead of $\psi$    we obtain that (\ref{eem2}) and  (\ref{wu})  are modified to
\begin{subequations}
\label{hr}
\begin{alignat}{4}
E^2(V)&=\phi(V-V_1)^2(V-V_3), \label{u22}\\
\frac{\phi}{4}&=2U^2LC_0\beta. \label{nn}
\end{alignat}
\end{subequations}
From   (\ref{hr})   we see that the soliton  exists both for positive and negative $\beta$. For positive $\beta$ we have dark soliton ($V_3<V_1$), and for negative $\beta$ -- bright soliton ($V_3>V_1$).

To find $V(Z)$ we have to solve Eq. (\ref{def}).
The solution  is
\begin{eqnarray}
\label{e59}
V(Z)=V_1+\left(V_3-V_1\right)
\sech^2\left(\frac{V_3-V_1}{2}\phi  Z\right).
\end{eqnarray}

\subsection{How good is the quasi-continuum approximation?}

We would like to discuss an opportunity (and motivation)
to go beyond the limits of
the quasi-continuum approximation used in the paper.
To do it, one may keep the third term in the Taylor expansion of the discrete second order derivative in the r.h.s. of (\ref{a8a}) \cite{kogan2,kogan3}
\begin{eqnarray}
 V_{n+1}-2V_{n}+V_{n-1} =\frac{\partial^2 V}{\partial Z^2}+\frac{1}{12}
\frac{\partial^4V}{\partial Z^4}+\frac{1}{360}
\frac{d^6V}{d Z^6}.
 \end{eqnarray}
 Thus we obtain
instead of (\ref{new4}) the equation
\begin{eqnarray}
\label{new22}
U^2L\frac{dQ}{d Z} =\frac{dV}{d Z}+\frac{1}{12}
\frac{d^3V}{d Z^3}+\frac{1}{360}
\frac{d^5V}{d Z^5}.
\end{eqnarray}
Looking at the results of the previous Sections we understand  that  the fifth order derivative in Eq. (\ref{new22}) can be  discarded if
$\gamma (V_1-V_2)^2/\alpha\ll 1$ and $\beta\left(V_1-V_3\right)/\alpha\ll 1$.
The small numerical coefficient before the sixth order derivative in the r.h.s. of (\ref{new22})  means that the small parameters
shouldn't be actually very small.

With all that we have to admit that the transition from the discrete to the continuous system is an uncontrollable approximation. More specifically, the mere fact of existence of the travelling wave of the form (\ref{comsi}) in the discrete system
should be proved analytically or, at least, numerically. We hope to return to this issue in the future.

\section{The dissipative kinks and shocks}
\label{lossy}

\begin{figure}[h]
\includegraphics[width=\columnwidth]{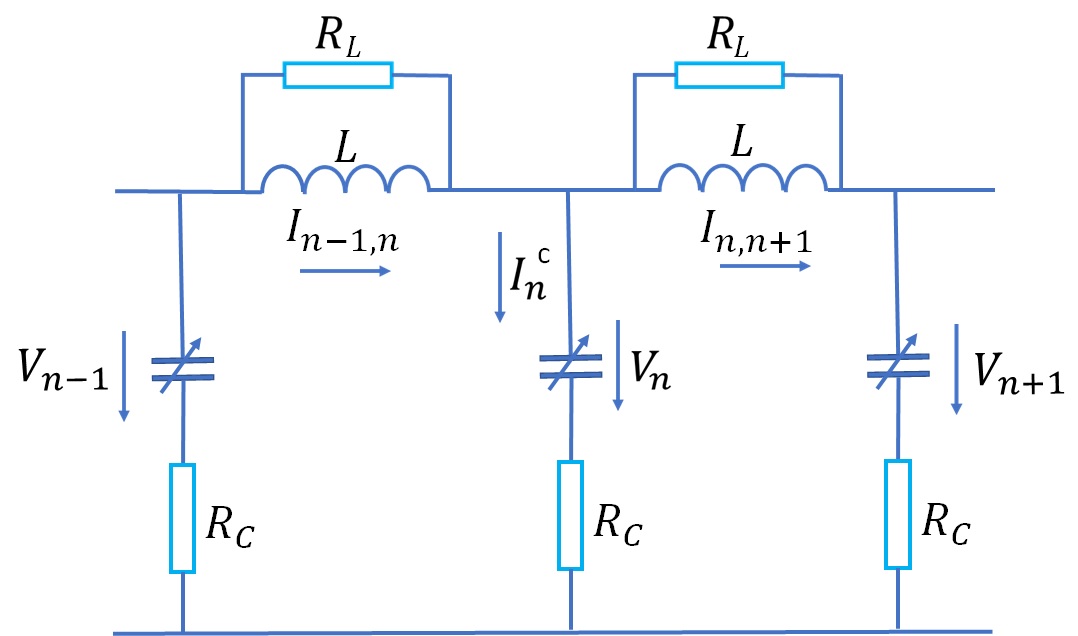}
\caption{The lossy discrete  transmission line.}
 \label{sh}
\end{figure}

Let us ask ourselves what happens with the kinks when the transmission line becomes lossy? To answer this question let us include  in the system the resistors $R_L$  shunting the inductors and the resistors $R_C$ in series with the nonlinear capacitors  (see  Fig. \ref{sh}).
In this case the circuit equations  are modified to
\begin{subequations}
\label{modi}
\begin{alignat}{4}
L\frac{d I_{n,n+1}}{d t}&=\left(1+R_CC\frac{d}{dt}\right)\left(V_{n}-V_{n+1}\right),
\label{n1a}\\
I^c_n&=\left(1+\frac{L}{R_L}\frac{d}{d t}\right)\left(I_{n-1,n}-I_{n,n+1}\right) \label{n1b}
\end{alignat}
(Eq. (\ref{a8b}) remains as it was).
Hence (\ref{comic}) is modified to
\end{subequations}
\begin{eqnarray}
\label{comic2}
L\frac{d^2 Q_n}{d t^2}=\left(1+\frac{L}{R_L}\frac{d}{d t}\right)\left(1+R_CC\frac{d}{dt}\right)\nonumber\\
\left(V_{n+1}-2V_{n}+V_{n-1}\right),
\end{eqnarray}
and  (\ref{com}) is modified due to the presence of ohmic resistors to
\begin{eqnarray}
\label{com3}
L\frac{\partial^2 Q(V)}{\partial t^2}=
\frac{\partial^2 V}{\partial Z^2}+\tau_{R}\frac{\partial^3V}{\partial t\partial^2Z}
+\frac{1}{12}\frac{\partial^4V}{\partial Z^4},
\end{eqnarray}
where
\begin{eqnarray}
\tau_R\equiv\frac{L}{R_L}+R_CC.
\end{eqnarray}
In deriving (\ref{com3}) we assumed that  the dissipation is small, so we ignored the terms which correspond to the product of the dissipation coefficients and the higher order derivatives with respect to $Z$.
For the travelling wave   we obtain from (\ref{com3})
after  integration the generalization of (\ref{new4})
\begin{eqnarray}
\label{new4b}
\frac{1}{12}\frac{d^3V}{d Z^3}+U\tau_{R}\frac{d^2V}{d Z^2}=U^2L\frac{dQ}{d Z} -\frac{dV}{d Z}.
\end{eqnarray}
Similar to what we said in Section \ref{con2}, notice that equation with the same form as (\ref{new4b}) one would obtain after making the travelling wave ansatz in the generalised  Korteweg-de Vries-modified Korteweg-de Vries (KDV-mKDV) equation with dissipation.

Performing  the second  integration, we obtain the generalization of (\ref{n45bu})
\begin{eqnarray}
\label{n45b}
\frac{1}{12}\frac{d^2V}{d Z^2}+U\tau_{R}\frac{dV}{d Z}=U^2LQ(V)-V+F.
\end{eqnarray}

From  (\ref{modi})  we obtain that in the present case the sum of the currents flowing through the nonlinear inductance and the ohmic resistance $R_L$ is given by the equation
\begin{eqnarray}
\label{i22}
I(Z)=-\frac{V(Z)}{UL}-\frac{\tau_R}{L}\frac{dV(Z)}{dZ}+\text{const}.
\end{eqnarray}

Equation (\ref{n45b}) is Newton equation,  describing  motion with friction of the fictitious particle  ($V$ being the coordinate of the particle and $Z$ - the "time") in the force field given by the r.h.s. of the equation.
We  again impose on $V(Z)$ the boundary conditions (\ref{vk}).
Hence the motion starts at $Z=-\infty$ at $V=V_1$ and ends at $Z=+\infty$ at $V=V_2$ (obviously that both $V_1$ and $V_2$ are the equilibrium positions).
Equating the r.h.s. of (\ref{n45b}) to zero at $V=V_1$ and $V=V_2$ we recover Eq. (\ref{ququ}) for the speed of the wave.

It will be convenient to present Eq. (\ref{n45b}) as
\begin{eqnarray}
\label{3b}
\frac{1}{12}\frac{d^2V}{d Z^2}+U\tau_{R}\frac{dV}{d Z} =-\frac{d\Pi(V)}{dV},
\end{eqnarray}
where
\begin{eqnarray}
\label{5bb}
\frac{d\Pi(V)}{dV}=-\frac{V_2-V_1}{Q_2-Q_1}Q(V)+V+\frac{V_2Q_1-V_1Q_2}{Q_2-Q_1}.
\end{eqnarray}
The constraint  for the travelling wave asymptotic boundary conditions is the inequality
\begin{eqnarray}
\label{bb}
\Pi(V_1)>\Pi(V_2).
\end{eqnarray}
Inequality (\ref{bb}) can be explicitly presented as
\begin{eqnarray}
\label{6b}
\int^{V_1}_{V_2}Q(V)dV<\frac{1}{2}\left(Q_1+Q_2\right)\left(V_1-V_2\right).
\end{eqnarray}
For  $\gamma>0,\beta\geq 0$, the condition (\ref{6b}) means just $|V_1|>|V_2|$.

The point $V=V_1$ should be the position of unstable equilibrium.
The point $V=V_2$ can be the point of either stable, or unstable equilibrium. (We met with similar dichotomy while studying localised travelling waves in Josephson transmission line \cite{kogan4}). The waves corresponding to the stable equilibrium at the point $V=V_2$  we called the shocks, to the unstable equilibrium -- the kinks.

Differentiating  (\ref{5bb}) we obtain the r.h.s. of (\ref{22m}) (see below)
 \begin{eqnarray}
\label{5bp}
-\frac{d^2\Pi(V)}{dV^2}=\frac{U^2}{u^2(V)}-1;
\end{eqnarray}
the point $V=V_i$ being the point of the stable/unstable equilibrium corresponds to the travelling wave being subsonic/supersonic with respect to the background.
Stable equilibrium at the point $V=V_2$ (together with unstable equilibrium
at the point $V=V_1$)  reflect the well-known in the nonlinear
waves theory fact: the shock speed is  larger than the sound
speed in the region before the shock but smaller than the sound
speed in the region behind the shock \cite{whitham}.

Let us return to (\ref{speeds}) giving the sound and
 the kink/shock  speeds for the cubic case.
The  kinks/shocks phase diagram in the space $V_1,V_2$   can be obtained  by comparing $U^2$ and $u^2(V_2)$.
Consider the case $\gamma>0,\beta\geq 0$.
If  $V_1$ and $V_2$ have the same sign, the travelling wave is always the shock.
In the opposite case, the wave can be either  the kink or the shock, depending upon $V_1$ and $V_2$.
The boundary in the phase space  between the kinks and the shocks  is especially simple for $\beta=0$
\begin{eqnarray}
V_1+2V_2=0
\end{eqnarray}
(very much similar to what we had for the Josephson transmission line \cite{kogan4}).

The most important distinction of the dissipative kinks from the shocks:
since the particle  stops at the unstable equilibrium point, for  the  kink to exist,
 the fine tuning   is necessary - the parameters $V_1$, $V_2$ and $R$ should satisfy definite relation (see below). The shocks don't need such find tuning -- they are generic.

To study quantitatively dissipative kinks and shocks
we will use the approach of the previous Section.
After  repeating exactly what was done there starting from Eq. (\ref{new4b}), we obtain modified
Eq. (\ref{22})
\begin{eqnarray}
\label{22m}
\frac{1}{24}\frac{d^2E^2}{dV^2}+U\tau_{R}\frac{dE}{dV}
=\frac{U^2}{u^2(V)}-1.
\end{eqnarray}
Considering again the cubic case we can easily check up that  (\ref{eem})
is still the solution of Eq. (\ref{22m}), with $\psi$ being still given by Eq. (\ref{velos}), but Eq. (\ref{k2}) is generalized to
\begin{eqnarray}
\label{2bb}
V_1+V_2&=-\frac{2\beta}{3\gamma}+\frac{4U\tau_R}{\psi}.
\end{eqnarray}
Hence the dissipative kinks and the shocks are of the same shape  (given by (\ref{kk})) as the kinks in the lossless line.

Now we understand what all the trouble of introducing the new dependent and independent variables was for. We were able to integrate analytically Eq. (\ref{22m}) (while we didn't see any way to integrate analytically Eq. (\ref{n45b})).

We need to emphasise the different meaning of the condition (\ref{2bb}) for the kinks and the shocks. For the kinks,  (\ref{2bb}) is the condition of the solution existence (and the solution happens to be  described by the elementary function). For the shock, (\ref{2bb}) is just the condition for the solution to be described by an elementary function. But, of course, the shock with any other values of $V_1,V_2,\tau_R$ also exists (provided that $V_1,V_2$ satisfy the constraint  (\ref{bb})).
So the obtained shock can be called the particular shock of the first kind.
Note  also that by integrating   (\ref{22m}) we can obtain Abel equation of the second kind, and (\ref{2bb}) is the condition of the solvability of the equation \cite{polyanin}.

Let us assume that $V_1$ and $V_2$ are so close to each other, that we can keep in the r.h.s. of (\ref{delo}) only the first two terms, thus simplifying (\ref{22m}) to
\begin{eqnarray}
\label{22bmb}
\frac{1}{24}\frac{d^2E^2}{dV^2}+U\tau_{R}\frac{dE}{dV}
=U^2LC_0(\alpha+2\beta V)-1
\end{eqnarray}
(we assume for the sake of definiteness that $\beta>0$ and $V_1,V_2>0$).
The elementary  solution of (\ref{22bmb})  satisfying the boundary conditions (\ref{vk7}) is of the form \cite{kogan4}
\begin{eqnarray}
 \label{an}
E=\chi \left(V-V_2\right)\left[\left(V-V_2\right)^{1/2}
-\left(V_1-V_2\right)^{1/2}\right].
\end{eqnarray}
Substituting (\ref{an}) into (\ref{22bmb})  we obtain
\begin{subequations}
\begin{alignat}{4}
\chi^2&=8U^2LC_0\beta\\
U\tau_R&=\frac{5\chi}{24}\left(V_1-V_2\right)^{1/2}.\label{hre}
\end{alignat}
\end{subequations}
Similar to  Eq. (\ref{2bb}) in the case of shocks,  Eq. (\ref{hre}) is just the condition for the shock to be expressed by an elementary function.
The shock exists for any other value of $V_1-V_2$.
So the obtained shock can be called the particular shock of the second kind.

The function $E(V)$ being found, we can find the function $V(Z)$ by solving Eq. (\ref{def}).
The solution is
\begin{eqnarray}
\label{fi8}
V(Z)=V_2+\frac{V_1-V_2}
{\left[\exp \left(\chi Z/2\right)+1\right]^2}.
\end{eqnarray}

\section{Conclusions}

We considered localised travelling waves of the voltage and the current in the discrete lossless transmission line with nonlinear capacitors. We have shown that these waves are of two kinds: the kinks, when the voltage has different asymptotic values
at both sides of the transmission line,
and the solitons (solitary waves), when the voltage has the same asymptotic value
at both sides of the transmission line.
  We also studied
lossy transmission lines. The  localised travelling waves in such systems are the dissipative kinks and the shocks.

The velocity of the kink, both in the lossless and in the lossy line, and of the shock  in the lossy line is given by the same equations
\begin{eqnarray}
\frac{1}{U^2}=L\frac{Q_2-Q_1}{V_2-V_1},
\end{eqnarray}
where $Q_1, V_1$ and $Q_2,V_2$ are the asymptotic values of the capacitor charge and the voltage on both ends of the transmission line.

To our opinion the most important equation of the paper is Eq. (\ref{22m}) describing the lossy transmission line
\begin{eqnarray}
\label{im}
\frac{1}{24}\frac{d^2E^2}{dV^2}+U\tau_{R}\frac{dE}{dV}
=\frac{U^2}{u^2(V)}-1.
\end{eqnarray}
When the charge of the non-linear capacitor and the voltage on it are connected by the equation
\begin{eqnarray}
Q(V)=\text{const}+C_0\left(\alpha V+\beta  V^2+\gamma V^3\right),
\end{eqnarray}
the profiles of the kink, both in the lossless and in the lossy line, and of the particular shock of the first kind in the lossy line, can be easily calculated on the basis of the solution of Eq. (\ref{im})
\begin{eqnarray}
E(V)=\psi(V-V_1)(V-V_2),
\end{eqnarray}
and are given by the same equation
\begin{eqnarray}
V(Z)=\frac{V_1+V_2}{2}-\frac{V_1-V_2}{2}
\tanh \left[\frac{V_1-V_2}{2}\psi  Z\right],
\end{eqnarray}
where
$\psi^2=6U^2LC_0\gamma$,
and the condition on the asymptotic values of the voltage on both ends of the transmission line $V_1$ and $V_2$ are  given by  Eq. (\ref{2bb}).
The profile of the particular shock of the second kind is given by Eq. (\ref{fi8}.

\begin{acknowledgments}

I am grateful to A. M. Kamchatnov,   B. Malomed and A. Vainchtein  for the discussion.
The paper was finalized during the Author' participation in Workshop on Twistronics and Moire Materials: Bridging Theory and Experiments, ICTP, Trieste. The Author is grateful to the Center for the hospitality.

\end{acknowledgments}


\begin{thebibliography}{99}

\bibitem{malomed2} E. Kengne, W. M. Liu, L. Q. English, and B. A. Malomed,
 Phys. Rep.  {\bf 982}, 1 (2022).

\bibitem{solitons} M. Remoissenet, {\it Waves Called Solitons: Concepts and Experiments},
Springer-Verlag Berlin Heidelberg GmbH (1996).

\bibitem{malomed4}
E. Kengne, R. Vaillancour, And B. A. Malomed,
Int. J.  Mod. Phys. B {\bf 23}, 133 (2009).

\bibitem{cameroon}
S. Abdoulkary, A. Mohamadou, and T. Beda,
Comm. Nonl. Sci Numer Sim. {\bf 16}, 3525 (2011).

\bibitem{bogdan}
M.M. Bogdan and D. V. Laptev, J.  Phys. Soc.  Japan {\bf 83}, 064007 (2014).

\bibitem{fermi} {\it he Fermi-Pasta-Ulam Problem: A Status Report (Ed: G. Gallavotti),
Lecture Notes in Physics}, Vol. 728, Springer, New York, NY 2007.

\bibitem{kevrekidis} P. G. Kevrekidis  and M. I. Weinstein, Physica D: Nonlinear Phenomena {\bf 142},  113 (2000).

\bibitem{vainchtein} A. Vainchtein,
 Physica D: Nonlinear Phenomena, {\bf  434},  133252 (2022).

\bibitem{vainchtein2} A. Vainchtein and Lev Truskinovsky, arXive:5376227;
Physica D: Nonlinear Phenomena 134187  (2024).

\bibitem{kamchatnov} L. F. Calazans de Brito  and A. M. Kamchatnov, Phys. Rev. E {\bf 109}, 015102 (2024).

\bibitem{kogan2}
E. Kogan,
Phys. Stat. Sol. (b) {\bf 259}, 2200160 (2022). https://doi: 10.1002/pssb.202200160

\bibitem{kogan3}
E. Kogan,
Phys. Stat. Sol. (b) {\bf 260}, 2200475 (2023). https://doi: 10.1002/pssb.202200475.


\bibitem{landauer1} R. Landauer, IBM J. Res. Dev. {\bf 4}, 391 (1960).

\bibitem{landauer2} S. T. Peng, R. Landauer, IBM J. Res. Dev.  {\bf 17}, 299 (1973).

\bibitem{landauer} R. Landauer,
J. Appl. Phys. {\bf 51} 5594  (1980).


\bibitem{kogan4}
E. Kogan,  Phys. Stat. Sol. (b), 2300336 (2024); https://doi.org/10.1002/pssb.202300336.

\bibitem{whitham} G. B. Whitham, {\it Linear and Nonlinear Waves}, John Wiley \& Sons Inc.,
New York (1999).

\bibitem{polyanin} A. D. Polyanin and V. F. Zaitsev, {\it Handbook of
exact solutions for ordinary differential equations, Second edition},
(Chapman \& Hall/CRC, Boca Raton London New York Washington, D.C.
2003).

\end{thebibliography}
\end{document}